\documentclass[final,1p,times]{elsarticle} 
\usepackage{graphicx} 
\usepackage{amssymb} 
\usepackage{amsthm} 

\journal{Nuclear Physics A} 
\begin{document} 

\begin{frontmatter} 


\title{Nonperturbative Heavy-Quark Interactions in the QGP}

\author{Ralf Rapp$^{a}$, Felix Riek$^{a}$, Hendrik van Hees$^{b}$,
Vincenzo Greco$^{c}$, Massimo Mannarelli$^{d}$}

\address[a]{Texas A\&M University, 
Cyclotron Institute and Physics Department, 
College Station, TX, 77843-3666, USA}
\address[b]{Justus-Liebig-Universit\"at Giessen, 
Heinrich-Buff-Ring 16, D-35392 Giessen, Germany}
\address[]{INFN-LNS, Laboratori Nazionali del Sud, 
and Dipartimento di Fisica e Astronomia, Universit\'a di Catania, Italy}
\address[b]{IEEC/CSIC, Universitat Aut\`onoma de Barcelona, 
Torre C5, E-08193 Bellaterra (Barcelona), Spain} 

\begin{abstract} 
We adopt a $T$-matrix approach to study quarkonium properties and 
heavy-quark transport in a Quark-Gluon Plasma. The $T$-matrix approach 
is well suited to implement potential scattering and thus provides a 
common framework for low-momentum transfer interactions in heavy-heavy 
and heavy-light quark systems. We assume that the underlying potentials 
can be estimated from the heavy-quark free energy computed in lattice 
QCD. We discuss constraints from vacuum spectroscopy, uncertainties 
arising from different choices of the potential, and the role of elastic 
and inelastic widths which are naturally accounted for in the $T$-matrix 
formalism. 
\end{abstract} 

\end{frontmatter} 


\section{Introduction}
\label{sec_intro}
Interactions of heavy quarks at low momentum transfer, $|q^2|\le m_Q^2$, 
are parametrically dominated by elastic scattering; Bremsstrahlung is 
suppressed by the large mass, $m_Q$, of the quark. Furthermore,  
3-momentum transfer dominates over energy transfer, 
$q_0\sim \vec q^2/2m_Q \ll |\vec q|$, which corresponds to the static 
limit or potential-type interactions. Both heavy-quark (HQ) transport 
and quarkonium binding are governed by low-momentum interactions. It is 
therefore suggestive to (i) address both problems in a common framework,
and (ii) invoke a potential-based description. A finite-temperature
$T$-matrix approach is well suited for these purposes; it provides 
a consistent framework to evaluate both bound-state and scattering 
solutions, based on a two-body static potential (see, e.g., 
Ref.~\cite{Rapp:2009my} for a recent review). An extra benefit arises
if the latter can be defined as a parameter-free input, e.g., from 
finite-temperature lattice QCD (lQCD), or at least be constrained by 
lQCD ``data".  In the regime of moderate or even strong coupling, 
resummations of large diagrams are necessary, which in the $T$-matrix 
equation is realized via the standard ladder sum.
In this paper, we set up a scattering equation with heavy quarks, 
check constraints in the zero-temperature and small-coupling 
(perturbative) limits and discuss the question of input potentials 
at finite temperature (Sec.~\ref{sec_tmat}), followed by an application 
to HQ diffusion and nonphotonic electron spectra (Sec.~\ref{sec_diff})
and conclusions (Sec.~\ref{sec_concl}).  

\section{$T$-Matrix Approach with Heavy Quarks}
\label{sec_tmat}
Starting from a relativistic Bethe-Salpeter equation, a 
3-dimensional (3D) Lippmann-Schwin\-ger equation can be derived by  
reducing the energy-transfer variable. A partial-wave expansion
then yields a 1D integral equation for the $T$-matrix, 
\begin{equation}
\label{Tmat}
T_{\alpha}(E;q',q)=V_{\alpha}(q',q) + \int \frac{2dk\,k^2}{\pi}
V_{\alpha}(q',k) \ G_{2}(E;k) \  
T_{\alpha}(E;k,q) \ [1-f_F(\omega_k^Q)-f_F(\omega_k^q)] \ , 
\end{equation}
in a given quantum-number channel $\alpha$. The concrete form of the 
intermediate 2-particle propagator, $G_{2}$, depends on the reduction 
scheme; applications to the the nuclear many-body problem or to 
quark scattering in the QGP~\cite{Mannarelli:2005pz} yield rather 
robust results.

\begin{figure}[!t]
\centering
\includegraphics[width=0.49\textwidth]{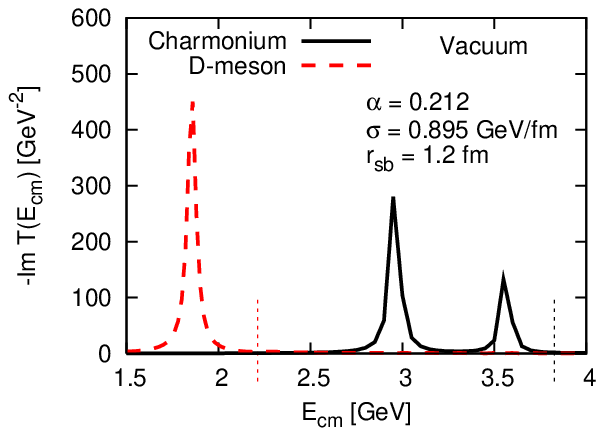}
\includegraphics[width=0.47\textwidth]{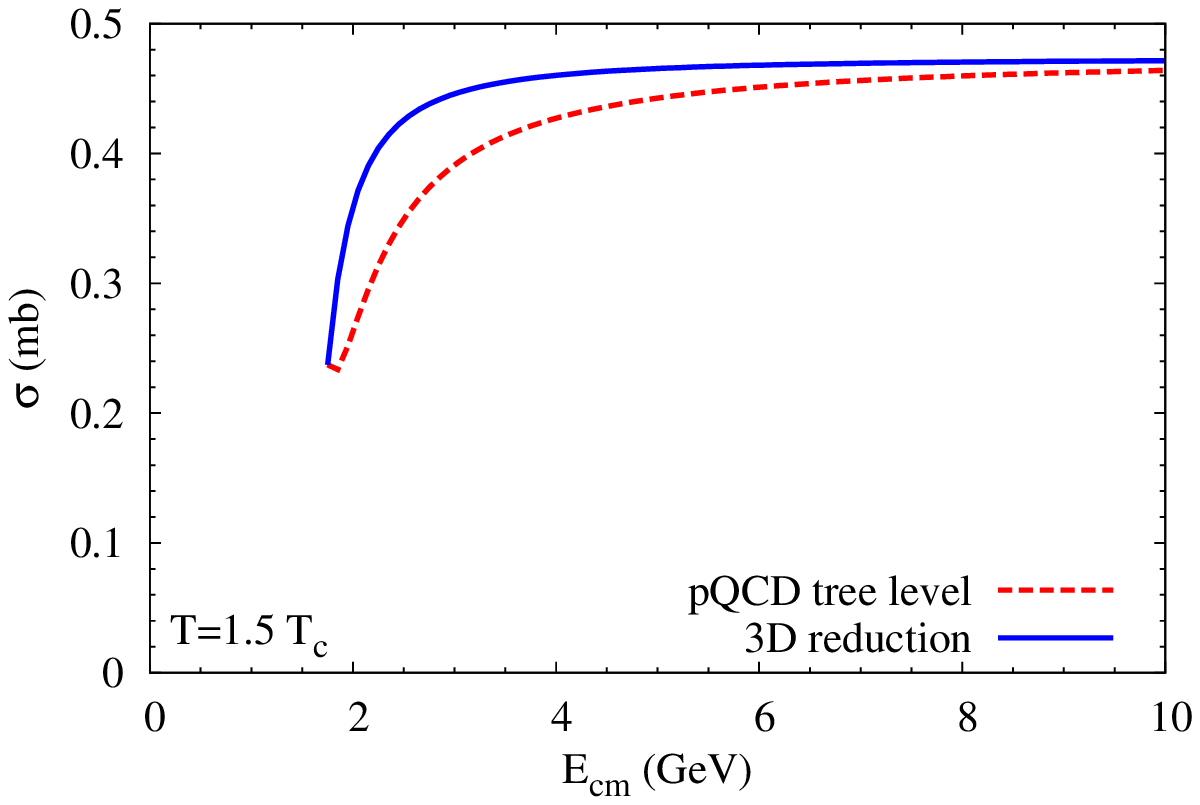}
\caption[]{Left: imaginary part of the $T$-matrix in vacuum for 
$S$-wave $c\bar q$ ($D$, $D^*$ mesons) and $c\bar c$ ($\eta_c$, 
$J/\psi$ mesons) channels using a $T$=0 HQ free (=internal) energy 
with string breaking at $r_{\rm sb}$=1.2~fm~\cite{Kaczmarek:2005ui}.
Right: comparison of the heavy-light quark cross section in LO 
pQCD with the Born term of a $T$-matrix calculation with
screenend Yukawa potential.}
\label{fig_vac-pqcd}
\end{figure}

In the HQ sector in vacuum, potential models have been established as 
the proper realization of low-energy QCD. The underlying potential is 
identified with the HQ free energy computed in lQCD (and agrees 
well with the phenomenological Cornell potential). As a first 
application, we inject this potential~\cite{Kaczmarek:2005ui} 
(subtracted at a typical string-breaking scale of $r_{\rm sb}$=1.2\,fm 
and Fourier-transformed) into Eq.~(\ref{Tmat}) to compute the vacuum 
spectrum in the $S$-wave $Q$-$\bar{Q}$ and $Q$-$\bar{q}$ channels. With 
bare HQ masses of $m_{c,b}^0$=1.4, 4.75\,GeV and a light-quark mass of 
$m_q$=0.35\,GeV, the $D$, $B$, $J/\psi$ and 
$\psi'$ as well as $\Upsilon$, $\Upsilon'$ and $\Upsilon''$ states are 
reproduced within an accuracy of $\sim$0.1\,GeV (spin-spin interactions, 
of order $1/m_Q$, are neglected here), cf.~left panel of 
Fig.~\ref{fig_vac-pqcd}. 

For applications to HQ spectra in heavy-ion collisions it is important
to check the high-energy and perturbative limit of the $T$-matrix. To 
this end we compare in the right panel of Fig.~\ref{fig_vac-pqcd}  
two calculations for the $c$-$\bar q$ cross section: 
(i) a leading-order (LO) perturbative QCD (pQCD) calculation 
with a screened one-gluon exchange propagator $1/(q^2-\mu_D^2)$; 
(ii) the Born approximation to the $T$-matrix using a Yukawa potential
(including a Breit correction to account for color-magnetic 
interactions~\cite{Brown:2003km}) 
with identical Debye-mass ($\mu_D$=$gT$) and strong coupling as in (i). 
The agreement is within $\sim$20\% and
shows that the Born term of the $T$-matrix is consistent with pQCD at 
high energies ($T(E_{\rm cm}) \to V(E_{\rm cm})$ for large $E_{\rm cm}$, 
see Ref.~\cite{Cabrera:2006wh}).
\begin{figure}[!t]
\centering
\includegraphics[width=0.48\textwidth]{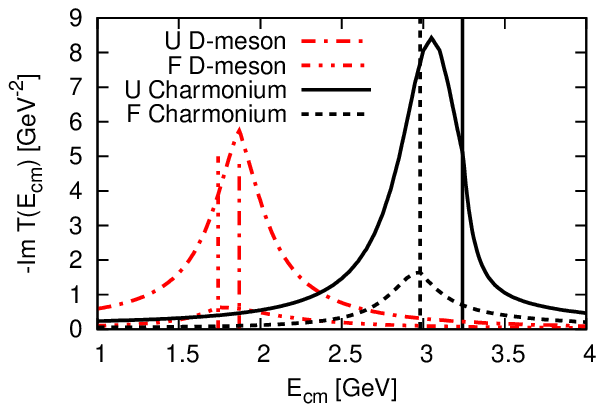}
\includegraphics[width=0.48\textwidth]{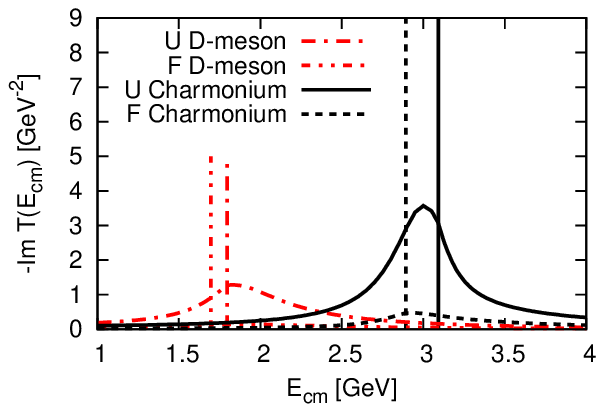}
\caption[]{Imaginary parts of the $c$-$\bar{q}$ and $c$-$\bar{c}$ 
$S$-wave $T$-matrices in the QGP at $T$=1.2\,$T_c$ (left) 
and 1.8\,$T_c$ (right) using potentials corresponding to the lQCD free 
(dashed lines) or internal (solid lines) energy~\cite{Petreczky:2004pz};
vertical lines indicate the quark-antiquark thresholds, 2$m_{Q}$ and 
$m_Q$+$m_q$; a single-quark width of 0.2~GeV is used in the 2-particle 
propagator, $G_2$.}
\label{fig_tmat}
\end{figure}

In Fig.~\ref{fig_tmat} we summarize our results for the in-medium 
$T$-matrices in the open and hidden charm sector. If the potential
is identified with the singlet free energy computed in thermal lQCD,
the effects are rather weak: at 1.2~$T_c$ the charmonium ground-state 
is about to become unbound while in the $D$-meson sector only a moderate 
threshold enhancement is visible; at 1.8~$T_c$ little strength is left 
even at threshold. However, if the corresponding internal energy is 
employed, the charmonium bound state is still (slightly) bound at 
1.2(1.8)\,$T_c$, while the $D$-meson channel exhibits an appreciable 
(moderate) ``Feshbach resonance" at threshold~\cite{vanHees:2007me}.

\section{Heavy-Quark Diffusion in QGP}
\label{sec_diff}
\begin{figure}[!t]
\begin{minipage}{0.5\linewidth}
\vspace{-0.4cm}
\includegraphics[width=0.9\textwidth]{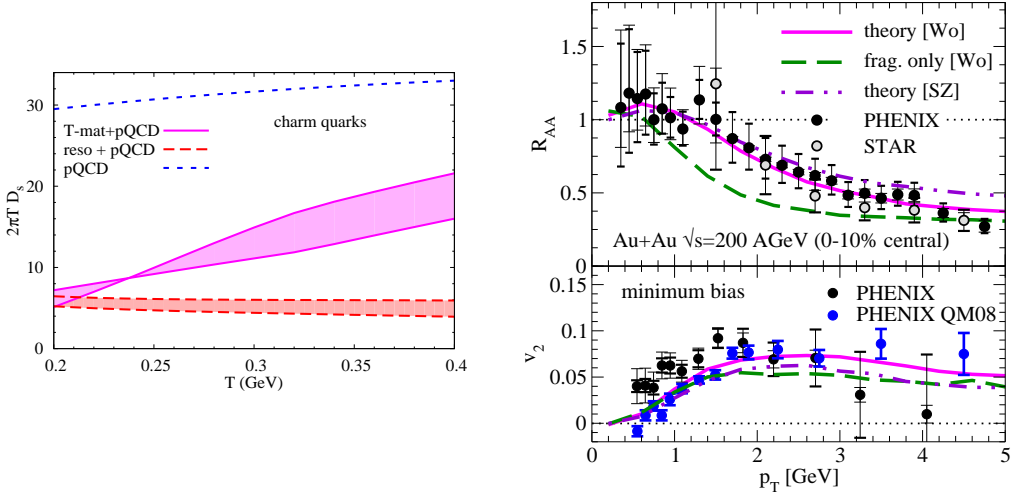}
\end{minipage}
\begin{minipage}{0.5\linewidth}
\includegraphics[width=0.95\textwidth]{RAA-v2-elec-tmat.eps}
\end{minipage}
\caption[]{Left: spatial $c$-quark diffusion constant in the $T$-matrix 
approach~\cite{vanHees:2007me}, effective resonance 
model~\cite{vanHees:2004gq} and LO pQCD 
($\alpha_s$=0.4)~\cite{Svetitsky:1987gq}. 
Right: comparison of semiletponic $e^\pm$ RHIC 
data~\cite{Adare:2006nq,Abelev:2006db} to relativistic Langevin 
simulations for $c$ and $b$ quarks using $T$-matrix+pQCD interactions
in an expanding QGP (including hadronization into $D$-/$B$-mesons and
subsequent 3-body decays); solid and dash-dotted lines are obtained 
with different lQCD internal energies; the dashed line 
is computed without coalescence contributions.  
}
\label{fig_diff}
\end{figure}

The coupling of heavy quarks to the expanding medium in heavy-ion 
collisions is a direct means to extract transport properties of the 
QGP. E.g., the degree of $c$-quark thermalization appears to be high
but not complete~\cite{Adare:2006nq,Abelev:2006db}, thus enabling a 
quantitative determination of the thermal relaxation time, $\tau_c$ 
(i.e., the latter is comparable to the QGP lifetime). The large HQ 
mass enables to compute transport coefficients within a Fokker-Planck 
equation. These are readily calculated using in-medium 
heavy-light quark $T$-matrices~\cite{vanHees:2007me}.
The resulting spatial diffusion coefficient (related to the 
thermal relaxation time by $D_s$=$\tau_Q T/m_Q$), is compared to 
other approaches in the left panel of Fig.~\ref{fig_diff}.     
Close to $T_c$, the value of $D_s$ in the $T$-matrix approach (with
internal energy as potential) is about a factor of 4-5 smaller than a
LO pQCD calculation (with $\alpha_s$=0.4), but comparable to
an effective $D$-meson resonance model~\cite{vanHees:2004gq}.
The increase of $D_s$ with temperature within the $T$-matrix 
calculations is due to a decreasing interaction strength caused by 
color-screening in the potential. It is a direct reflection of the
dissolving $D$-resonance structures (recall Fig.~\ref{fig_tmat});
toward high temperatures the pQCD results are approached.
We note that the use of free energy as potential leads to a diffusion
constant similar to the LO pQCD calculations.   

The temperature- and momentum-dependent diffusion coefficients have
been implemented into Langevin simulations of $c$ and $b$ quarks
in Au-Au collisions at RHIC~\cite{vanHees:2005wb,vanHees:2007me}. The 
elliptically expanding medium has been modeled by a thermal fireball 
in close resemblance of hydrodynamic simulations~\cite{Kolb:2003dz}.
A QGP evolution with initial temperature $T_0$$\simeq$340\,MeV is 
followed by a QGP-hadron-gas mixed phase with a constant total entropy 
to reproduce the observed number of charged hadrons. The final HQ 
spectra are hadronized in a combined coalescence/fragmentation scheme 
and decayed into electrons to compare to experiment, cf.~right panel of 
Fig.~\ref{fig_diff}. Note the role of quark coalescence processes at $T_c$ 
in increasing both the $R_{AA}$ and $v_2$. Coalescence may be viewed as a 
manifestation of the $T$-matrix interaction in the 
hadronization process.  

\section{Conclusions}
\label{sec_concl}
A $T$-matrix approach for elastic HQ scattering in the QGP has been used 
to study open and hidden heavy flavor in a common framework. Assuming 
interaction potentials given by the HQ internal energy extracted from 
thermal lQCD, HQ thermalization is substantially accelerated compared 
to pQCD estimates. This is caused by resonant correlations in heavy-light 
quark scattering at temperatures up to $\sim$1.5~$T_c$. Charmonia remain 
bound up to $\sim$2\,$T_c$. Much weaker effects emerge when the 
HQ free energy is employed. 


\section*{Acknowledgments} 
This work has been supported by a
U.S. National Science Foundation CAREER award under grant
no. PHY-0449489 and by the A.-v.-Humboldt foundation through a Bessel
award (RR).


\begin{thebibliography}{00} 

\bibitem{Rapp:2009my}
  R.~Rapp and H.~van Hees,
  to be published in {\it QGP-4} (R.C.~Hwa and X.-N.~Wang, eds.); 
  arXiv:0903.1096 [hep-ph]

\bibitem{Mannarelli:2005pz}
  M.~Mannarelli and R.~Rapp,
  Phys.\ Rev.\  C {\bf 72} (2005) 064905.

\bibitem{Kaczmarek:2005ui}
  O.~Kaczmarek and F.~Zantow,
  Phys.\ Rev.\  D {\bf 71} (2005) 114510.
                                                                                
\bibitem{Petreczky:2004pz}
  P.~Petreczky and K.~Petrov,
  Phys.\ Rev.\  D {\bf 70} (2004) 054503.

\bibitem{Brown:2003km}
  G.E.~Brown, C.H.~Lee, M.~Rho and E.~Shuryak,
  Nucl.\ Phys.\  A {\bf 740} (2004) 171.

\bibitem{Cabrera:2006wh}
  D.~Cabrera and R.~Rapp,
  Phys.\ Rev.\  D {\bf 76} (2007) 114506.

\bibitem{vanHees:2007me}
  H.~van Hees, M.~Mannarelli, V.~Greco and R.~Rapp,
  Phys.\ Rev.\ Lett.\  {\bf 100} (2008) 192301.
   
\bibitem{vanHees:2004gq}
H.~van Hees and R.~Rapp,
Phys.\ Rev.\  C {\bf 71} (2005) 034907.

\bibitem{Svetitsky:1987gq}
  B.~Svetitsky,
  Phys.\ Rev.\  D {\bf 37} (1988) 2484.

\bibitem{Adare:2006nq}
A.~Adare {et~al.} [PHENIX Collaboration], {Phys. Rev. Lett.}
  \textbf{98} (2007) 172301;  
T.C.~Awes  [PHENIX Collaboration],
  J.\ Phys.\ G {\bf 35} (2008) 104007.

\bibitem{Abelev:2006db}
B.I. Abelev {et~al.} (STAR Collaboration), {Phys. Rev. Lett.}
  \textbf{98} (2007) 192301.

\bibitem{vanHees:2005wb}
  H.~van Hees, V.~Greco and R.~Rapp,
  Phys.\ Rev.\  C {\bf 73} (2006) 034913.

\bibitem{Kolb:2003dz}
P.F.~Kolb and U.W.~Heinz  (2003), published in R.C.~Hwa and X.-N.~Wang
(editors), \textit{Quark-gluon plasma} vol. 3 (World Scientific, 2004)
p.~634, {arXiv:nucl-th/0305084}.


\end{thebibliography}
\end{document}